\documentclass[12pt]{article}
\usepackage{color}
\usepackage{amsmath} %Never write a paper without using amsmath for its many new commands 
\usepackage{amssymb} %Some extra symbols 
\usepackage[english]{babel}
\usepackage{graphics}
\usepackage[pdftex]{graphicx}
\usepackage{xspace}
\usepackage[gen]{eurosym}
\usepackage[font=footnotesize,format=plain,labelfont=bf,up,textfont=it,up]{caption}
\usepackage[left]{lineno}
\usepackage{sectsty}
\sectionfont{\large}
\subsectionfont{\small}
\begin{document}

%\linenumbers

%\date{}
%\includegraphics[width=0.15\textwidth]{indico_logo.pdf}
%\vskip -30pt \hspace{70pt}
%\noindent{\em Contribution to the European Strategy for Particle Physics --}

%\vskip -3pt\hspace{60pt} {\em  Open Symposium Preparatory Group, Kracow 10-12 September 2012}
%\title{\bf Search for anomalies in the neutrino sector with muon spectrometers and large LArÐTPC imaging detectors at CERN}
%\label{sec:future_exps_lar_nessie}

%\nopagebreak[4]

%\maketitle

\vskip 20pt
\begin{center}
{\bf\Large  An Appraisal of Muon Neutrino Disappearance \\ at Short Baseline}
%\vskip 7pt
%{\bf\Large with muon spectrometers}
\vskip 10pt
L.~Stanco$^1$, S.~Dusini$^1$, A.~Longhin$^2$, A.~Bertolin$^1$, M.~Laveder$^3$
\end{center}

\noindent {\em \footnotesize $^1$INFN-Padova, Via Marzolo, 8, I-35131 Padova, Italy, \\
 $^2$Laboratori Nazionali di Frascati, INFN, Via E. Fermi 40, I-00044 Frascati, Italy, \\ 
 $^3$Padova University and  INFN-Padova,  Via Marzolo, 8, I-35131 Padova, Italy}

\vskip 20pt

\vskip 20pt
\centerline{{\bf Abstract}}

\vskip 20pt

\begin{center}
\begin{minipage}{5in}
Neutrino physics is nowadays receiving more and more attention as a possible source of information for the long--standing problem
of new physics beyond the Standard Model. The recent measurement of the third mixing angle $\theta_{13}$ in the standard 
mixing oscillation scenario encourages us to pursue the still missing results on leptonic CP violation and absolute neutrino
masses. However, several puzzling measurements exist which deserve an exhaustive evaluation.
We will illustrate the present status of the muon disappearance measurements at small $L/E$ and the current CERN project
to revitalize the neutrino field in Europe with emphasis on the search for sterile neutrinos. We will then illustrate the achievements that a double muon 
spectrometer can make with regards to discovery of new neutrino states, using a newly developed analysis.
\end{minipage}
\end{center}

\vskip 200pt

{\em To be published in ``Advances in High Energy Physics''.}

\newpage

\section{Introduction}

The unfolding of the physics of the neutrino is a long and exciting story spanning the last 80 years. Over this time the interchange of 
theoretical hypotheses and experimental facts has been one of the most fruitful demonstrations of the progress of knowledge in physics.
The work of the last decade and a half finally brought a coherent picture within the Standard Model (SM) (or some small extensions of it),
namely the mixing of three neutrino flavour states with three  $\nu_1$, $\nu_2$ and $\nu_3$ mass eigenstates. 
The last unknown mixing angle, $\theta_{13}$, was recently measured~\cite{theta13-DB, theta13-RE, theta13-DC, theta13-T2} but still
many questions remain unanswered to completely settle the scenario: the absolute masses, 
the Majorana/Dirac nature and the existence and magnitude of leptonic CP violation.
Answers to these questions will beautifully complete the (standard) three--neutrino model but they will hardly provide an insight into new physics 
Beyond the Standard Model (BSM). 
Many relevant questions will stay open: the reason for neutrinos, the relation between the
leptonic and hadronic sectors of the SM, the origin of Dark Matter and, overall, where and how to look for BSM physics.
Neutrinos may be an excellent source of BSM physics and their story is supporting that at length.

There are actually several experimental hints for deviations from the ``coherent'' picture described above.
Many unexpected results, not statistically significant on a single basis, appeared also in the last decade and a half,
bringing attention to the hypothesis of {\em sterile neutrinos}~\cite{pontecorvo}. A recent White Paper~\cite{whitepaper} contains a comprehensive
review of these issues. In this paper we will focus on one of the most intriguing and long--standing unresolved result:
the unexpected oscillation of neutrinos at relatively small values of the ratio $L/E$ (distance in km, energy in GeV), corresponding to a scale
of $\mathcal{O}$(1) eV$^2$, incompatible 
with the much smaller values related to the atmospheric $|\Delta m_{32}^2|\simeq 2.4 \times 10^{-3}$~eV$^2$ and to the solar 
$\Delta m_{21}^2 \simeq 8 \times 10^{-5}$ eV$^2$ scales.

The first unexpected measurement came from an excess of $\overline\nu_e$ originating from an initial $\overline\nu_\mu$ beam from
Decay At Rest (LNSD~\cite{lsnd}). The LSND experiment saw a 3.8 $\sigma$ effect.  The subsequent  experiment with $\nu_{\mu}$ ($\overline\nu_\mu$) 
beam from accelerator, 
MiniBooNE~\cite{larnessie_5}, although confirming an independent 3.8 $\sigma$ effect after sustained experimental work,
was unable to draw conclusive results on the origin of the LSND effect having observed an excess at higher $L/E$ in an energy region where background
is high.

In recent years many phenomenological studies were performed by analyzing the LSND effect together with similar
unexpected results  coming from the measurement of lower than expected rates  of $\overline{\nu}_e$ and $\nu_e$ interactions
({\em disappearance}), either  from {\bf (a)} near-by nuclear reactors ($\overline\nu_e$)~\cite{reattori} or {\bf (b)} from Mega-Curie K-capture calibration
sources in the solar $\nu_e$ Gallium experiments~\cite{larnessie_4}. These $\nu_e$ ($\overline\nu_e$) disappearance measurements, all at the statistical
level of 3-4 $\sigma$, could also be interpreted~\cite{giunti-laveder} as oscillations between neutrinos at large $\Delta m^2\simeq 1$~eV$^2$.
Several attempts were then tried to reach a coherent picture in terms of mixing between active and
sterile neutrinos, in $3+1$ and $3+2$~\cite{larnessie-6}  or even $3+1+1$~\cite{treunouno} or $3+3$~\cite{tretre} models, 
as extensions of the standard three--neutrino model.
We refer to~\cite{models,models1,giuntinew} as the most recent and industrious works where a very crucial issue is raised: 
``{\em a consistent interpretation of the global data in terms of neutrino oscillations is challenged by
the non-observation of a positive signal in $\nu_{\mu}$ disappearance experiments}''~\cite{models1}. In fact, in any of the above models, essential information 
comes from the disappearance channels ($\nu_{\mu}$ or $\nu_e$), which is one of the cleanest channels to measure the oscillation parameters 
(see Sect.~\ref{sect-3} for details).

The presence of additional sterile states introduces quite naturally appearance and disappearance phenomena involving the flavour states in
all channels. In particular, a $\nu_\mu$ disappearance effect has to be present and possibly
measured. It turns out that only old experiments and measurements are available for Charged Current (CC) $\nu_\mu$  
interactions at small $L/E$~\cite{CDHS}. The CDHS experiment reported in 1984 the non--observation of $\nu_\mu$ oscillations 
in the $\Delta m^2$ range $0.3$ eV$^2$ to $90$ eV$^2$. Their analyzed region of oscillation did not span however low values 
of the mixing parameter down to around $0.1$ in $\sin^2(2\theta)$. More recent results are available on $\nu_\mu$ disappearance
from MiniBooNE~\cite{mini-mu}, a joint MiniBooNE/SciBooNE analysis~\cite{mini-sci-mu1, mini-sci-mu2} and the Long--Baseline MINOS 
experiment~\cite{minos}. These results slightly extend the $\nu_\mu$ disappearance exclusion region, however
still leaving out the small--mixing region. Similar additional constraints on $\nu_{\mu}$ disappearance could possibly come
from the analysis of atmospheric neutrinos in IceCube~\cite{icecube}.

Despite this set of measurements being rather unsatisfactory when compared with the corresponding LSND allowed region 
that lies at somewhat lower values of the mixing angle, they are still sufficient to introduce 
tensions in all the phenomenological models developed so far~(see e.g. \cite{whitepaper,models,models1,giuntinew,reviews} for comprehensive and recent reviews).
Therefore it is mandatory to setup a new experiment able to improve
 the small--mixing angle region exclusion by at least one order of magnitude with respect to the current results. 
 In such a way one could also rule out the idea that the mixing angle extracted from LSND is larger than the true value due to a data over--fluctuation.
 Once again, the $\nu_\mu$ disappearance 
channel should be the optimal one to perform a full disentangling of the mechanism given the strong tension between the
$\nu_e$ appearance and $\nu_\mu$ disappearance around $\Delta m^2\simeq 1$ eV$^2$.
In fact, whereas the LSND effect may be confirmed by a more accurate $\nu_e$ oscillation measurement, only the  presence of a $\nu_\mu$ 
oscillation pattern could shed more light on the nature and the interpretation of the effect.

Further, in the paper we will briefly discuss the newly proposed CERN experiment, following a detailed analysis of possible
ways to measure the $\nu_\mu$ disappearance channel. In particular the evaluation of the  $\nu_\mu$
disappearance rates at two different sites leads us naturally to take their ratio, elucidating the possibility
to observe depletions or/and excesses. 
Throughout the paper we will focus on the need for a new  $\nu_\mu$ CC measurement corresponding to an increase of one order of magnitude
in the sensitivity to the mixing parameter, exploiting the determination of the muon charge 
for a proper evaluation of the different models and the separation of $\nu_{\mu}$ and $\overline{\nu}_\mu$ in the same neutrino beam.
Finally, we will pay particular attention to the consistency and robustness (in statistical terms) of the results.

\newpage
\section{The CERN experimental proposal}

The need for a definitive clarification on the possible existence of a neutrino mass scale around 1 eV has brought up several
proposals and experimental suggestions exploiting the sterile neutrino option by using different interaction channels and refurbished experiments.
In the light of the considerations discussed in the previous section there are essentially two sets of experiments which must be redone:
a)~the measurement of $\overline\nu_{e}$ neutrino fluxes at reactors (primarily the ILL one) together with refined and detailed computations
of the flux simulations
(see e.g. ~\cite{huber}); b)~the appearance/disappearance oscillation measurements at low $L/E$ with a standard muon neutrino beam with its intrinsic 
electron neutrino component.

There is actually another interesting option which comes from the Neutrino Factory studies and the very recently submitted 
EOI from $\nu$STORM~\cite{nustorm}. We refer to~\cite{winter} for a comprehensive review of the corresponding possible $\nu_e$ and $\nu_{\mu}$ 
disappearance effects. It is interesting to note that our figures of merit about $\nu_{\mu}$ disappearance are rather similar to or even slightly more
competitive than those illustrated in~\cite{winter} (e.g compare the exclusion regions in Fig.~6 of \cite{winter} and those in Fig.~\ref{ster-5} of this paper 
despite the use of different C.L.), not
forgetting the rather long time needed to setup the $\nu$STORM project.

Coming to experiments with standard beams, investigations are underway at CERN where two Physics Proposals~\cite{nessie, icarus} were submitted 
in October 2011 and later merged into a single Technical Proposal 
(ICARUS-NESSiE,~\cite{larnessie}). CERN has subsequently set up working groups for the 
proton beam extraction from the SPS, the secondary beam line and the needed infrastructure/buildings for the detectors. The work was reported in a 
recent LOI~\cite{edms}. 

The experiment is based on two identical Liquid Argon (LAr)--Time Projection Chambers (TPC)~\cite{icarus} complemented by magnetized 
spectrometers~\cite{nessie} detecting electron and muon neutrino events at far and near
positions, 1600~m and 460 m away from the proton target, respectively.
The project will exploit the ICARUS T600 detector, the largest 
LAr-TPC ever built of about 600 ton mass, now presently in the LNGS underground 
laboratory where it was exposed to the CNGS beam. It is supposed to be moved at the CERN ÒfarÓ position. 
An additional 1/4 of the T600 detector (T150) would be constructed from scratch as a clone of the original one, except for the dimensions,
 and located in the near site. Two spectrometers would be placed 
downstream of the two LAr-TPC detectors to greatly enhance the physics reach.
The spectrometers will exploit a bipolar magnet with instrumented iron slabs, and a newly designed
air--core magnet, to perform charge identification and muon momentum measurements in an extended energy range (from 0.5 GeV
or less to 10 GeV), over a transverse area larger than 50 m$^2$.

While the LAr-TPCs will mainly perform a direct measurement of electron neutrinos~\cite{rubbia} the spectrometers will
allow an extended exploitation of the muon neutrino component, with neutrino/antineutrino discrimination on an
event-by-event basis.  

\subsection{The neutrino beam}
The proposed new neutrino beam will be constructed in the SPS North Area~\cite{edms}. The setup is based on a 100 GeV proton beam with a fast extraction 
scheme providing about 3.5~$\cdot$~10$^{13}$ protons/pulse in two pulses of 10.5~$\mu$s durations\footnote{Pulses of 10.5~$\mu$s duration are normally  
put in coincidence
with the fast response of the spectrometers' detectors and efficiently used to reject the cosmic ray background. See e.g. ~\cite{opera-time} where a 
time resolution of less than 2 ns is reported for the detectors used in the OPERA experiment.} 
separated by 50~ms for a sample of about 4.5~$\cdot$~10$^{19}$ protons on target (p.o.t.) per year.
A target station will be located next to the so called  TCC2 target zone, 11~m underground, followed by a cylindrical He-filled decay pipe with a
length of about 110~m and a diameter of 3~m. The beam dump of 15~m in length, will be composed of iron blocks with a graphite inner
core. Downstream of the beam dump a set of muon chambers stations will act as beam monitors. The beam will point upward, with a slope of
about 5~mrad, resulting in a depth of 3 m for the detectors in the far site.

The current design of the focusing optics includes a pair of pulsed magnetic horns operated at relatively low currents. A graphite target
of about 1~m in length is deeply inserted into the first horn allowing a large acceptance for the focusing of low momentum pions emitted at
large angles. This design allows production of a spectrum peaking at about 2~GeV thus matching the most interesting domain of $\Delta
m^2$ with the detector locations at 460 and 1600~m from the target.

\begin{figure}[htbp]
\begin{center}
  \includegraphics[width=0.53\textwidth]{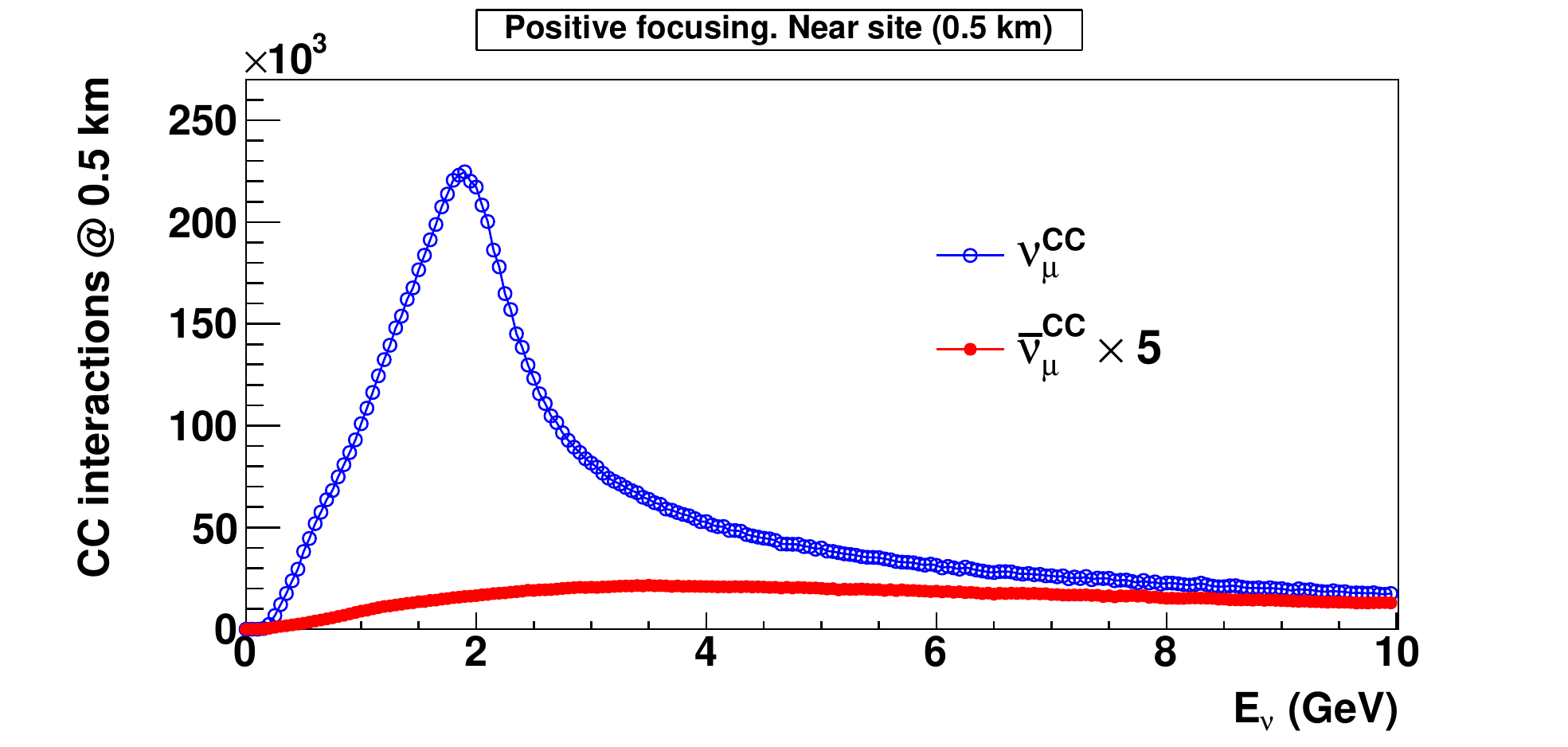}%
    \includegraphics[width=0.53\textwidth]{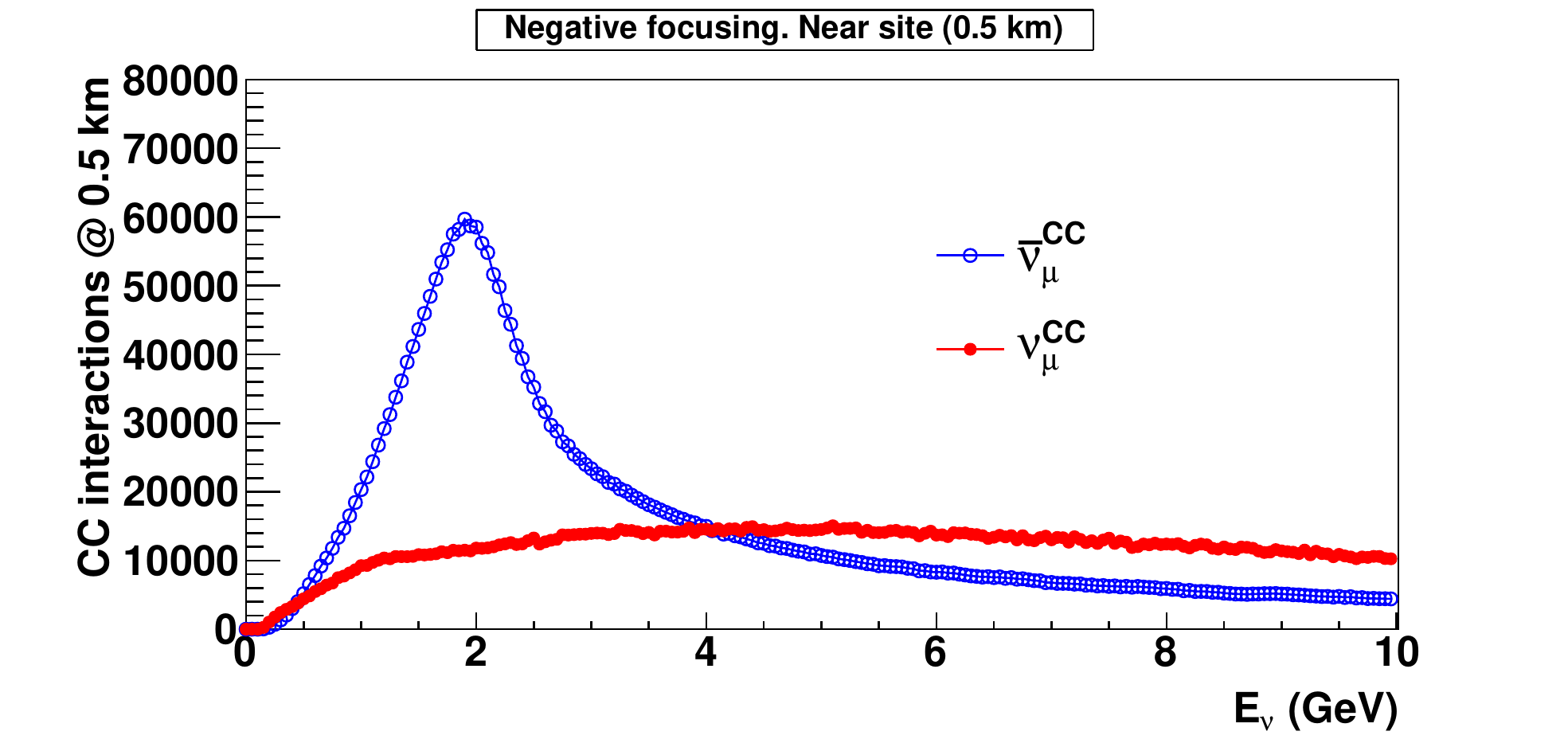}
  \includegraphics[width=0.53\textwidth]{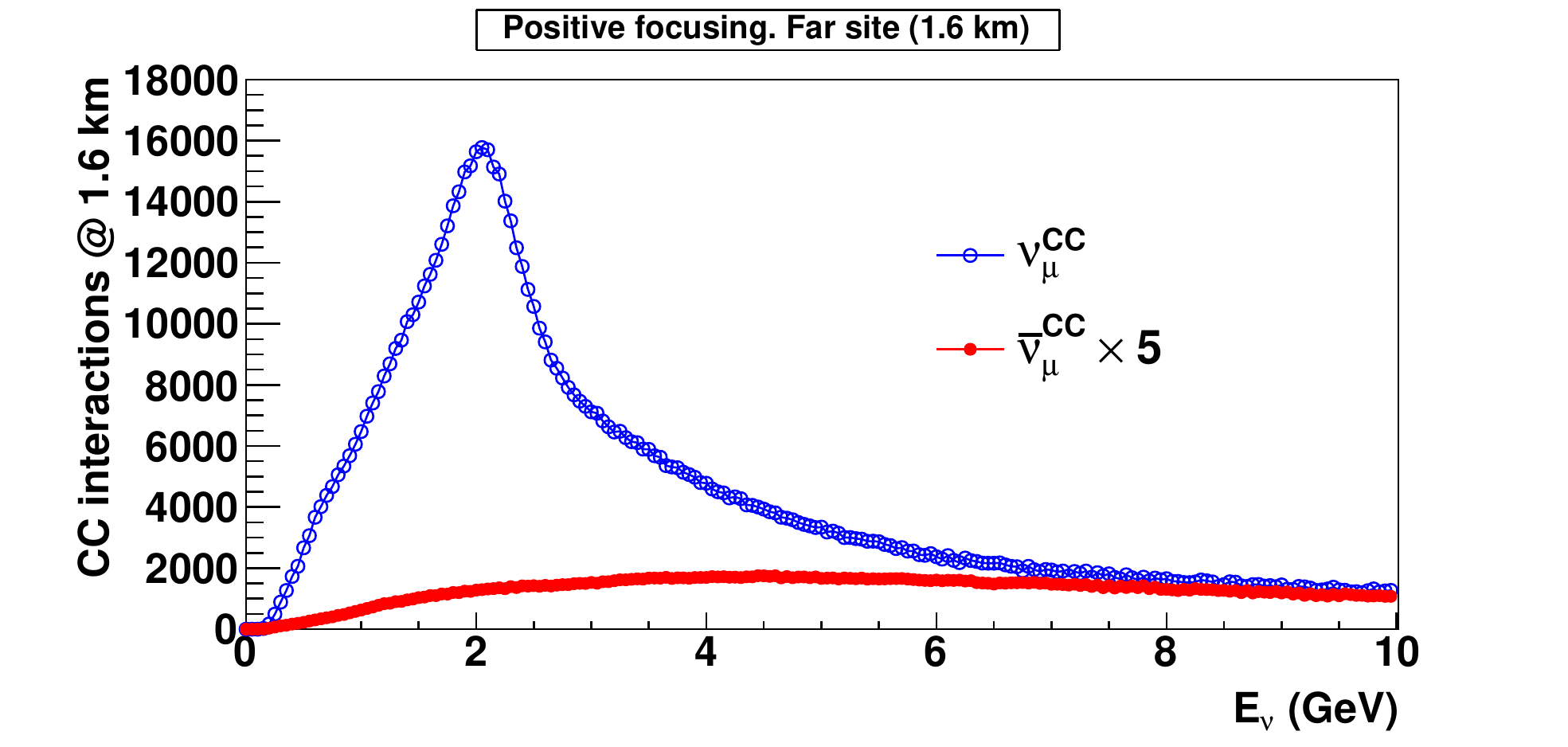}%
    \includegraphics[width=0.53\textwidth]{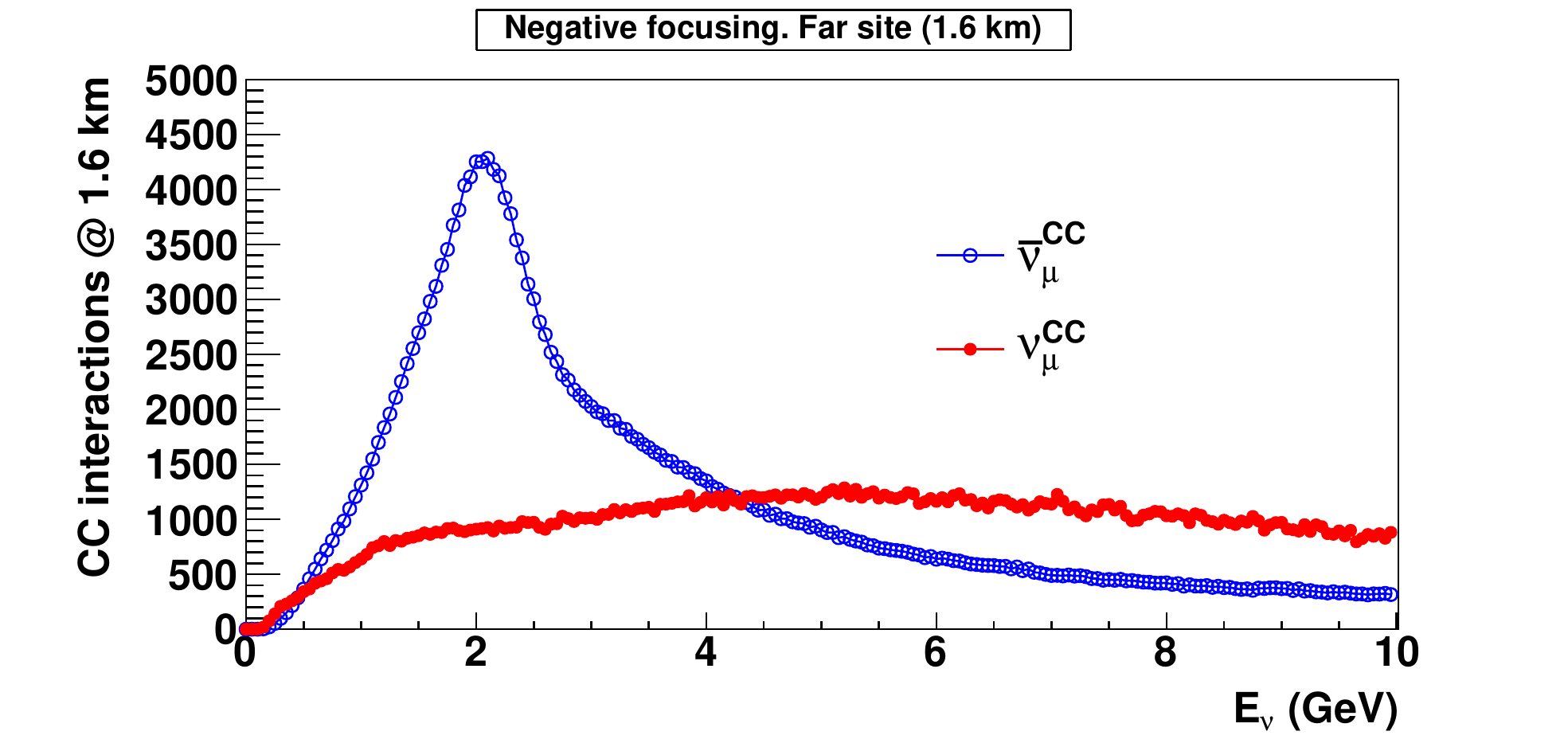}
  \includegraphics[width=0.53\textwidth]{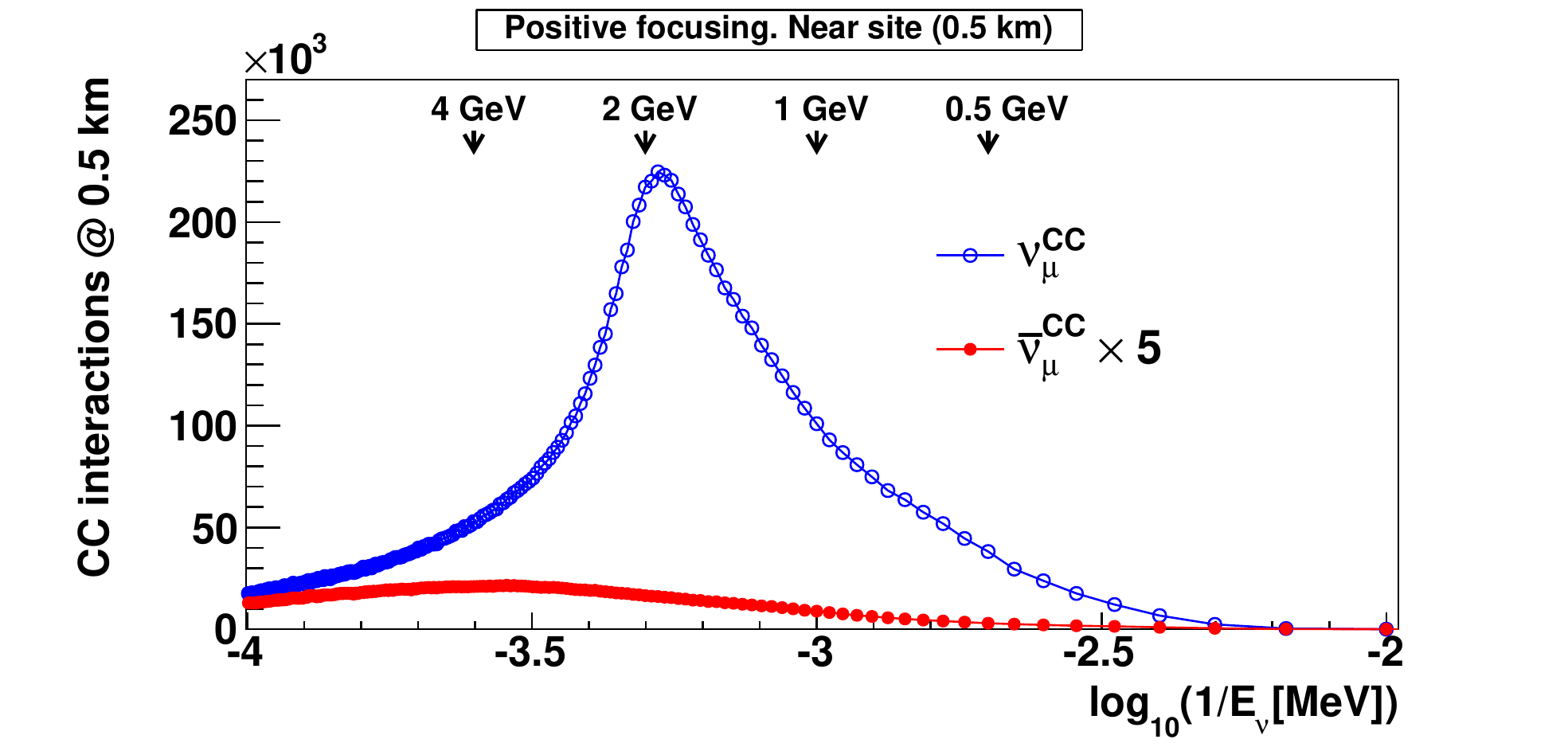}%
    \includegraphics[width=0.53\textwidth]{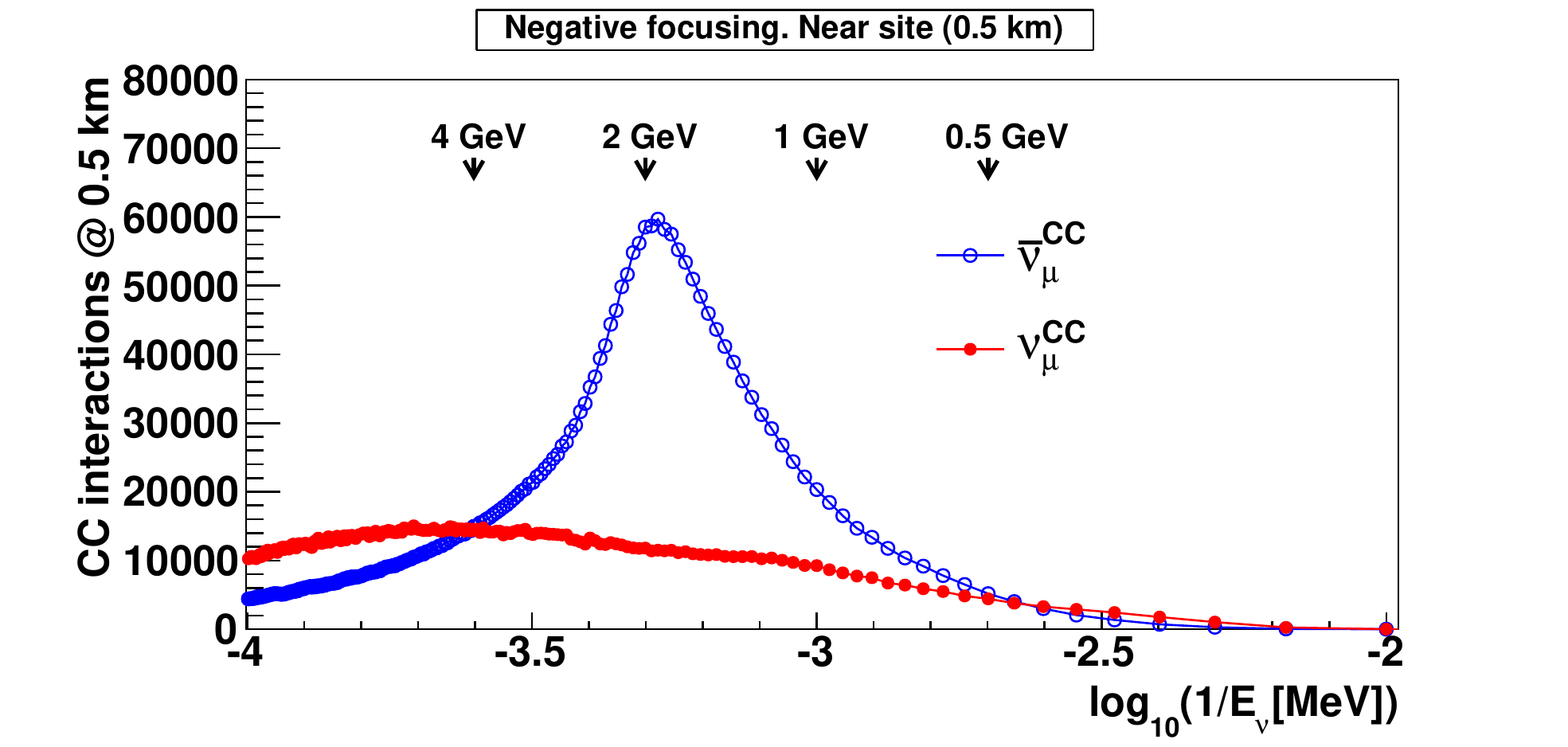}
    \caption{Expected neutrino CC interactions in the no--oscillation hypothesis for positive polarity (left) and negative polarity (right)
    for the new proposed CERN neutrino beam (elaborated from~\cite{edms}) and for an integrated luminosity of 1 year. 
    In the first (second) row the rates are shown at the near (far) position. In the third row the rates expected at the near site are displayed
     as function of $\log_{10}(1/E)$, used in the paper (see Sect.~\ref{sect-3}).
     Note that the distributions of the antineutrino rates in case of positive polarity are multiplied by a factor 5 to allow a better visual inspection.}
    \label{beam-AL}
\end{center}
\end{figure}

The charged current event rates for $\nu_\mu$ and $\bar{\nu}_\mu$ at the near and far detectors are shown in Fig.~\ref{beam-AL} for the positive and
negative focusing configuration.

A relevant contamination of $\nu_\mu$ in the negative polarity configuration is visible especially at high energy. This
component arises as a result of the decays of high energy poorly de--focused mesons produced at small angles. 
The charge discrimination of the magnetic system described below will allow an efficient discrimination of these two
components with a charge confusion below or of the order of 1\% from sub--GeV (0.3--0.5 GeV) up to momenta around 8--10 GeV~\cite{nessie}.

\subsection{Spectrometer requirements}

The main purpose of the spectrometers placed downstream of the LAr-TPC is to provide charge identification and momentum 
reconstruction for the muons produced in neutrino interactions occurring in the LAr volume or in the magnetized iron of the spectrometers. 
In order to perform this measurement with high precision and in a wide energy range, from sub-GeV to multi-GeV, a massive iron-core dipole magnet 
(ICM) is coupled to an air--core magnet (ACM) in front of it~\cite{nessie}. Low momentum muons will be measured by the ACM while the ICM will be 
employed at higher momenta. 

\begin{figure}[htbp]
\begin{center}
  \includegraphics[width=0.7\textwidth]{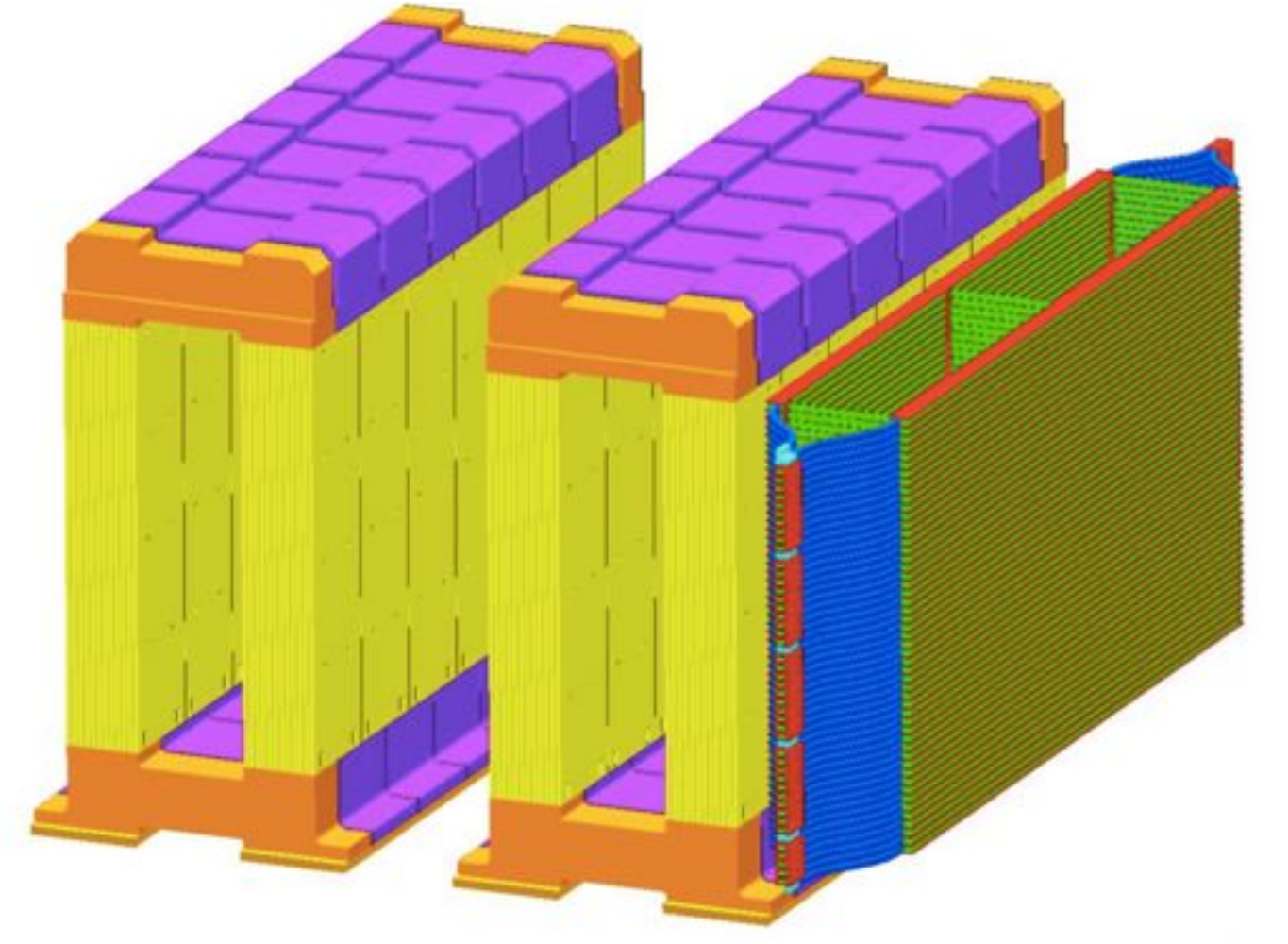}
    \caption{Sketch of the far site spectrometer that could be built by extensively reusing materials available from the OPERA 
     spectrometers~\cite{opera}. The basic elements of the new ACM concept are also depicted (courtesy of the NESSiE Collaboration).
     Neutrinos are traveling from right to left.}
    \label{nessie-far}
\end{center}
\end{figure}

As considered in the previous sections the definition of two sites, near and far, constitutes a fundamental issue for each physics program
which aims to perform any sterile neutrino search. The two layouts have to be as similar as possible in order to allow an
almost complete cancellation of the systematic uncertainties when comparing the measurements made at the near and far sites. 
Hence the near spectrometer will be an exact clone of the far one, with identical thickness along the beam axis but a scaled transverse size.
A sketch of the possible far site NESSiE detector is shown in Fig.~\ref{nessie-far}.

The key feature of the ACM is the large geometric and momentum acceptance. The need of a low momentum threshold for muon reconstruction can be
met using a magnet in air. The only dead material along the muon path is given by the conductors needed to generate the magnetic field and the
position detectors instrumenting the magnet itself. For the conductors the use of aluminum instead of copper is preferable
due to the lower $Z$ and a density lower by a factor 3. The magnetic field needed in the low momentum
range covered by the ACM is in the range 0.1 - 0.15 T. A spatial resolution in the range of 0.1 - 1 mm can be reached using drift tubes~\cite{opera-hpt} as
high precision trackers in combination with scintillator strip detectors~\cite{opera-scint}. These could provide the external trigger 
needed by the drift tubes and a coarse spatial measurement in the non--bending direction i.e. the direction parallel to the drift tubes. Silicon 
photomultiplier (SiPM) devices may eventually be used to read out some scintillator planes embedded in the magnetic field.

The general layout of the two OPERA~\cite{opera} iron spectrometers fulfills the requirements of the ICM detector, and
they could eventually be used for the CERN project putting the two magnets one after the other in order to obtain a total of 4 m longitudinal thickness of iron.

The OPERA spectrometers are built assembling vertical iron plates (slabs) in a planar structure of 875 cm (width) $\times$ 800 cm (height). Each 
passive plane is made out of seven adjacent iron slabs. Resistive Plate Chambers (RPC)~\cite{opera-rpc} are sandwiched between iron planes; 
21 RPC detectors, 
arranged in seven rows and three columns, are used in each active plane. Each RPC detector has a rectangular shape and covers an area of about
3.2 m$^2$. The RPCs provide tracking measurements with about 1 cm resolution using the digital read-out of strips with a 2.6 cm pitch in the bending direction
and 3.5 cm in the non--bending direction. The magnet is made of two arms with 22 RPC layers alternating with iron
layers. Top and bottom iron yokes are connecting the arms. Copper coils surrounding the yokes are used to generate a magnetic field of about 1.5 T in the iron 
circuit. 
A total of 924 RPC chambers  are needed to instrument each spectrometer. 
They could  be recovered from the OPERA spectrometers and re-used. 
These RPC chambers are standard 2 mm gap chambers with bakelite electrodes with resistivity in the range from 
10$^{11}$ to 5$\times$10$^{12}$ $\Omega$ $\cdot$ cm at $T=20 ^{\circ}$C. In OPERA they are operated in the streamer regime with a gas mixture made of 
$Ar/C_2H_2F_4/I$-$C_4H_{10}/SF_6$ in the volume ratios of 75.4/20/4/0.6 at five refills/day. The high-amplitude streamer signals produced by charged tracks crossing the
gas volume allow to house the Front-End discriminators in racks placed on top of the spectrometer.

\section{The $\nu_\mu$ disappearance analysis}\label{sect-3}

The use of at least two sites where neutrino interactions can occur is mandatory to observe the oscillation pattern, as established by almost all the most
recent neutrino experiments. Indeed the disappearance probability due to an additional sterile neutrino is given by the usual two-flavour formula:
$$ P(\nu_\alpha\rightarrow\nu_\alpha)=1-\sin^2(2\theta)\cdot\sin^2(1.267\cdot\Delta m^2\cdot L/E), \, \text{\small ($L$ in km, $E$ in GeV)}$$
where the disappearance of flavour $\alpha$ is due to the oscillation of neutrino mass states at the $\Delta m^2$ scale
and at an effective mixing angle $\theta$ that can be simply parametrized as a function of the elements of a $3+1$ extended mixing matrix. 
As $L$ is fixed by the experiment location, the oscillation is naturally driven by the neutrino 
energy, with an {\em amplitude} determined by the mixing parameter.

%Whereas the disappearance channels show amplitudes which depend only on the mixing/coupling between the observed flavour 
%and the unknown sterile(s) neutrino state(s), the appearance channels corresponds to a much more complicated picture with amplitudes
%which mix up the couplings between the initial and final flavour states and the sterile states. For example, neglecting quadratic terms in the mixing matrix 
%elements, the effective amplitude for $\nu_e$ appearance in a $\nu_{\mu}$ beam  is given by:
%$$\sin^2(2\theta_{\mu e})\approx\frac{1}{4}\sin^2(2\theta_{ee})\sin^2(2\theta_{\mu\mu})$$
%while the amplitude for the disappearance of flavor $\alpha$ depends only on the mixing matrix element $U_{\alpha4}$ as:
%$$A_{\alpha\alpha}=4|U_{\alpha4}|^2(1-|U_{\alpha4}|^2)$$
%in the $3+1$ framework. The latter property is kept in extended $3+n$ models.
%Therefore, in this respect only the measurement of the disappearance channels is directly sensitive to a marginalized coupling to sterile neutrinos.

The disappearance of muon neutrinos due to the presence of an additional sterile state
depends only on terms of the extended PMNS~\cite{pmns} mixing matrix ($U_{\alpha i}$ with $\alpha= e,\mu,\tau$ and $i=1$,\ldots,4)
involving the $\nu_{\mu}$ flavor state and the additional mass eigenstate 4. In a 3+1 model at Short Baseline (SBL) we have:
$$ P(\nu_{\mu}\to\nu_{\mu})_{SBL}^{3+1} = 1 - 4 \vert U_{\mu 4} \vert^2 (1 - \vert U_{\mu 4} \vert^2) \sin^2 \frac{\Delta m^2_{41} L}{4E}$$
In contrast, appearance channels (i.e. $\nu_\mu \to \nu_e$) are driven by
terms that mix up the couplings between the initial and final flavour states and
the sterile state yielding a more complex picture:
$$ P(\nu_{\mu}\to\nu_e)_{SBL}^{3+1} = 4 \vert U_{\mu 4}\vert^2 \vert U_{e 4} \vert^2  \sin^2 \frac{\Delta m^2_{41} L}{4E}$$
This also holds in extended $3 + n$ models.

%Therefore while the appearance channels depend on the interplay between appearance and disappearance cases, at the near and far sites,
%due to specific models of oscillations, in the disappearance channels the oscillation probabilities can be described by the same probability 
%at two different sites, near and far since the same single flavor is measured (more extensive discussions on this issue can be found e.g. in Sect. 2 of \cite{winter}).
%

It is interesting to notice that the appearance channel is suppressed by two more
powers in $\vert U_{\alpha 4}\vert$. Furthermore, since $\nu_e$ or $\nu_\mu$ appearance
requires $\vert U_{e 4}\vert > 0$ and $\vert U_{\mu 4}\vert > 0$, it should be naturally accompanied by
a corresponding $\nu_e$ and $\nu_\mu$ disappearance. In this sense the disappearance
searches are essential for providing severe constraints  on the models of the theory
(a more extensive discussion on this issue can be found e.g. in Sect. 2 of \cite{winter}).

It must also be noted that the number of $\nu_e$ neutrinos depends on
the $\nu_e\rightarrow\nu_s$ disappearance and $\nu_\mu\rightarrow\nu_e$ appearance, and, naturally,
from the intrinsic $\nu_e$ contamination in the beam. 
On the other hand, the amount of $\nu_\mu$ neutrinos depends only on the
$\nu_\mu\rightarrow\nu_s$ disappearance and $\nu_e\rightarrow\nu_\mu$ appearance but the latter is
much smaller due to the fact that the $\nu_e$ contamination in $\nu_\mu$ beams is
usually at the percent level. 
Therefore in the $\nu_{\mu}$ disappearance channel the oscillation probabilities in both near and far detectors can be measured without any interplay 
of different flavours, i.e. by the same probability amplitude.

\begin{figure}[htbp]
\begin{center}
  \includegraphics[width=0.60\textwidth]{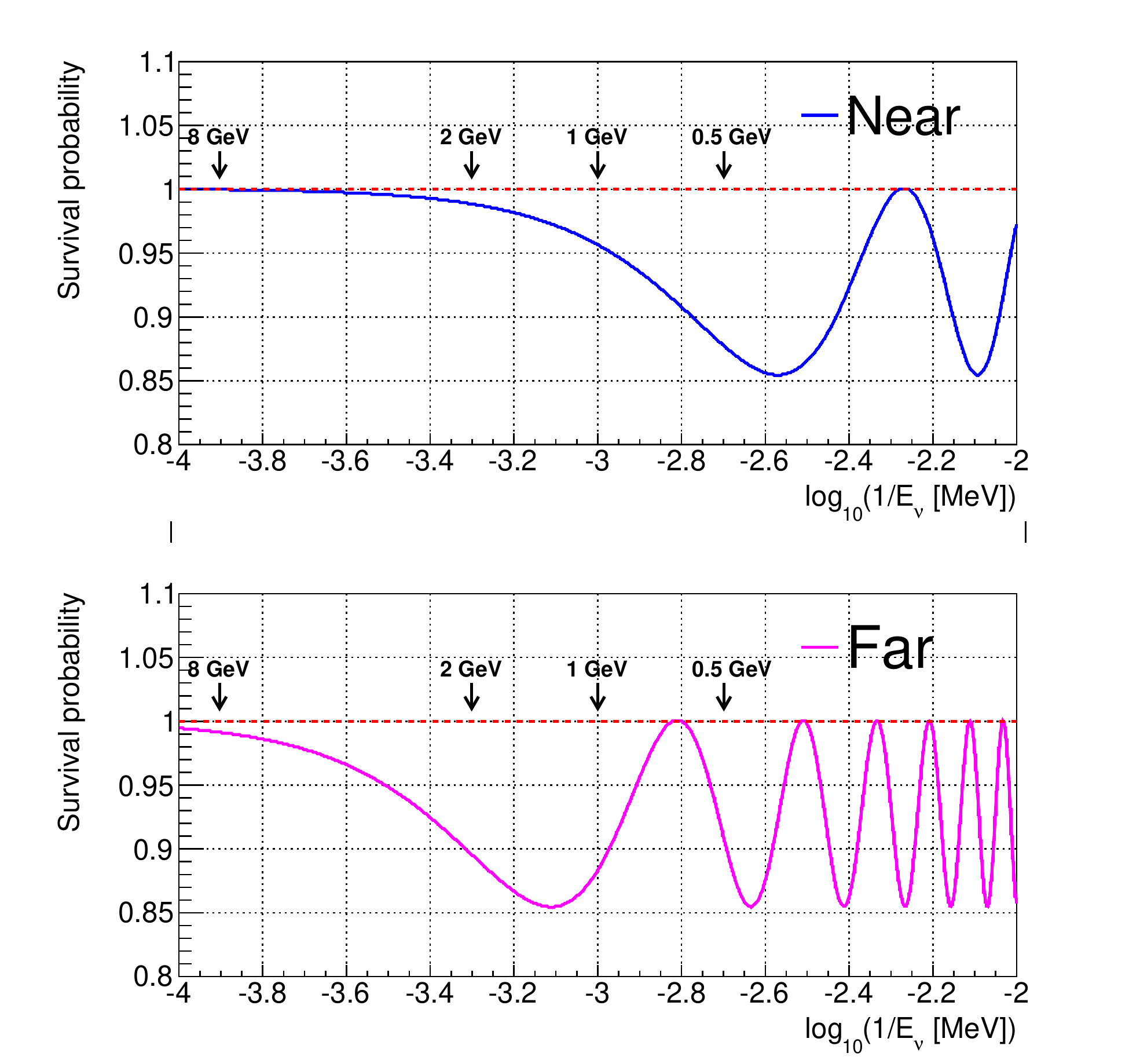}
    \caption{Disappearance probabilities in the two--flavour limit at near (top) and far (bottom) sites, at 460 and 1600 m, respectively, by using the amplitude 
    provided by the reactor anomaly, 0.146, and the mass scale $\Delta m^2=1$ eV$^2$. The x-axis corresponds to $\log_{10}(1/E_\nu)$,
    with $E$ in MeV.}
    \label{ster-1}
\end{center}
\end{figure}

\begin{figure}[htbp]
\begin{center}
  \includegraphics[width=0.7\textwidth]{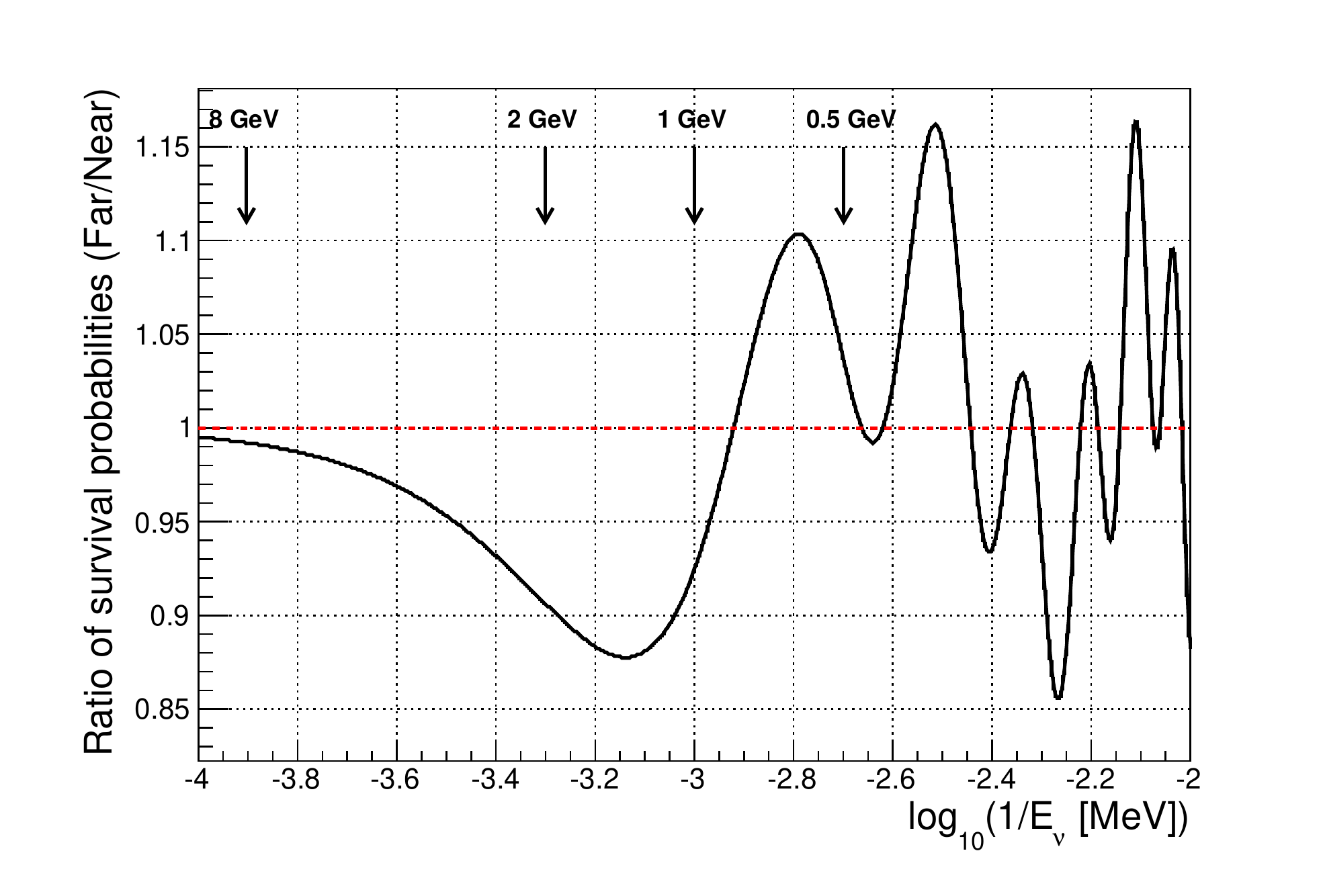}
    \caption{The ratio of the disappearance probabilities at near and far sites, 460 and 1600 m, respectively, in the two--flavour limit, by using the amplitude 
    provided by the reactor anomaly, 0.146, and the mass scale $\Delta m^2=1$ eV$^2$. The x-axis corresponds to $\log_{10}(1/E_\nu)$,
    with $E$ in MeV.}
    \label{ster-2}
\end{center}
\end{figure}

By taking into account the lengths $L$ of the locations given in the CERN proposal, {\em $L_{Near}=$} 460 m and {\em $L_{Far}=$}1600 m, we may plot 
the disappearance
probability as a function of the more {\em convenient} variable $\log_{10}(1/E)$, using $E$ in MeV to avoid singularities.
As an example, in Fig.~\ref{ster-1} the disappearance probability is shown for the near and far sites, by assuming an amplitude\footnote{It is
worthwhile to note that the disappearance amplitude affects the sensitivity to the mixing angle that in turn depends mainly on the statistical extent
of the data. Therefore the same evidence can be obtained for smaller amplitudes,
i.e. smaller effective mixing angles, by increasing the data sample. Finally the ultimate limit of an experiment
is set by the intrinsic systematic errors corresponding to the kinematics of the neutrino interaction and to the muon reconstruction sensitivity. 
Fig.~\ref{ster-6} below illustrates the issue.}
that corresponds to the averaged reactor anomaly $\overline\nu_e$ disappearance,  \mbox{$R = 0.927\pm 0.023$}~\cite{reattori}, and 
$\Delta m^2=1$ eV$^2$. It appears that at $\Delta m^2\approx 1$ eV$^2$ oscillations are already visible
at the near site. Such a behaviour may turn out to help the actual measurement by plotting 
the ratio between the far and near sites data as a function of $\log_{10}(1/E)$ (Fig.~\ref{ster-2}). The {\em Far/Near} ratio of the oscillation probabilities well illustrates the pattern that
could be observed at the CERN Short--Baseline experiment: a decrease above 1 GeV and an increase below 1 GeV of the neutrino 
energy! 

The disappearance patterns have to be evaluated after folding  cross-sections and efficiencies with the neutrino fluxes and detector geometry.
The {\em Far/Near} pattern does not depend at first order either on cross-sections and efficiencies or on the fluxes as far as they are
similar at the two sites, and within the use of two {\em } identical detector systems.
However, due to the different $\nu$ and $\overline{\nu}$ cross-sections, errors are enhanced when neutrinos and anti-neutrinos are not identified and separated on an event-by-event basis. Therefore the use of magnetic detectors is mandatory to keep systematics under control.
Then the largest source of error consists of the smearing  introduced by the reconstruction procedure.
Eventually this issue will be addressed by considering a conservative systematic effect in the muon momentum reconstruction.

\begin{figure}[htbp]
\begin{center}
  \includegraphics[width=0.8\textwidth]{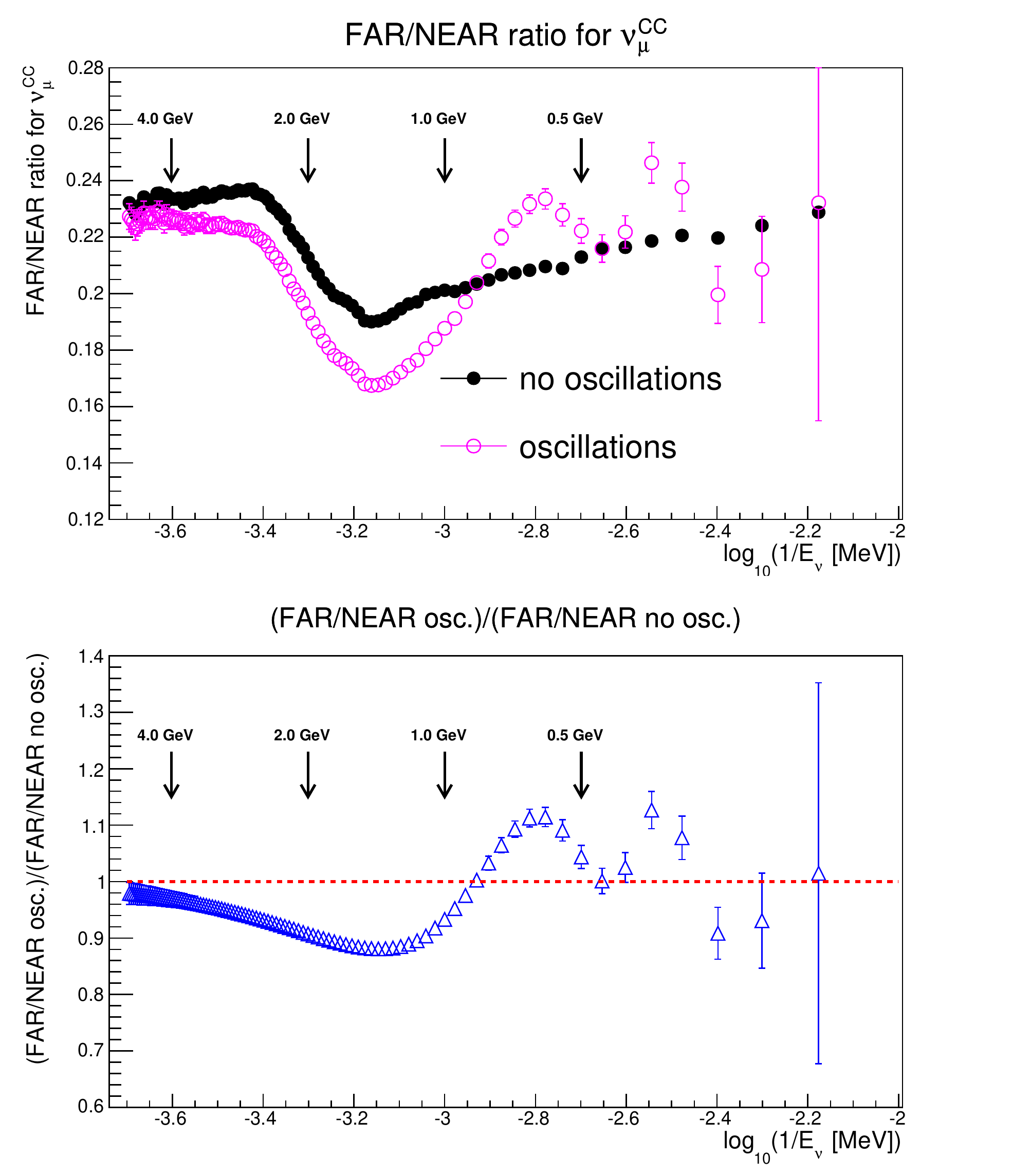}
    \caption{The oscillated and unoscillated $\nu_{\mu}$ event distributions, parametrized in $\log_{10}(1/E_{\nu})$, 
    for a luminosity of $4.5\times 10^{19}$ p.o.t. of the CERN $\nu_{\mu}$ 
    beam flux over the two magnet system described in the text, are plotted with their statistical errors. 
    On the top the {\em Far/Near} collected ratios are drawn.
    The no--oscillation shape is due to the convolution of the neutrino cross-section and the fluxes at the near and far site, while the
    oscillated shape is due to the sterile inclusion (with $\Delta m^2=1$ eV$^2$ and mixing amplitude equal to $0.146$). On the bottom
    the double ratio {\em (Far/Near)$_{oscillated}$/(Far/Near)$_{unoscillated}$} is shown for the same sterile assumption.}
    \label{ster-3}
\end{center}
\end{figure}

Fig.~\ref{ster-3} (top) illustrates a realistic observation of the non--oscillation probability when convoluted with
the estimated CERN $\nu_{\mu}$ beam flux and the neutrino cross--sections, by using the
detector systems described above and one year of data taking. The flatness is lost as expected. However a specific case of
oscillation is drawn, confirming that the workable behaviour is still present.
The correlated estimator, the double ratio parameter, defined as 
{\em (Far/Near)$_{oscillated}$/ (Far/Near)$_{unoscillated}$}, is also shown in the bottom of the Figure. 
It can be used for the statistical analysis.  
In the following we will demonstrate how the oscillation behaviour of the {\em Far/Near} parameter will allow an increase 
of an order of magnitude in the sensitivity to the mixing angle, taking into account the systematic errors, too.

Another parameter can be used in case of measurement of Neutral Current (NC) events become available, either with the proposed LAr detectors
or even with a less refined and massive detector. The NC/CC ratio becomes then a valuable estimator~\cite{nessie} that may enter 
into the evaluation of the significance of the sterile neutrino observation. A study that includes the information from Neutral Current events
is out of the scope of the present paper.
However for completeness we would like to briefly outline here the issue as it was originally described in~\cite{nessie}.

The observation of a depletion in NC events would be a direct and manifest signature of the existence of sterile neutrinos.
In fact the NC event rates are unaffected by standard neutrino mixing being flavour--blind such that their disappearance could only be explained by 
$\nu_{\mu}$ ($\nu_e$) $\rightarrow\nu_s$ transitions.
Even if NC events, either from $\nu_e$ or $\nu_{\mu}$, are efficiently detected by a Liquid Argon detector or a less refined detector, 
the transition rate measured with NC events has to agree with the 
$\nu_{\mu}$ CC  disappearance rate once the 
$\nu_{\mu}\rightarrow\nu_{\tau}$ and $\nu_{\mu}\rightarrow\nu_e$ 
contributions have been subtracted.
Therefore the NC disappearance is measured at best by the double ratio:
$({\frac{NC}{CC})_{Far}}/({\frac{NC}{CC})_{Near}}$.
The double ratio is the most robust experimental quantity to detect NC disappearance, once
$CC_{Near}$ and $CC_{Far}$ are precisely measured thanks to the spectrometers,
at the near and far locations. For that it is mandatory to  
disentangle $\nu_{\mu}$ and $\overline\nu_{\mu}$ contributions. 

\section{Simulation and results}\label{sect-4}

The magnetic detector system that has to be developed for the $\nu_{\mu}$ disappearance measurement should take into account all the 
considerations depicted in previous sections. The system developed by the NESSiE Collaboration~\cite{nessie} is actually well suited since it 
couples a very powerful high-$Z$ magnet for the momentum measurement via {range} to a low-$Z$ magnet, to extend the useful muon momentum interval as low as possible,
to allow charge discrimination on an event--by--event basis and to allow NC event measurement whether coupled to an adequate (but not necessarily
highly performant and with large mass) detector to identify NC events.

In the following we concentrate on the results achievable by the {\em Far/Near} estimators described in the previous section, by exploiting a full neutrino beam 
simulation, an up--to--date neutrino interaction model (GENIE 2.6~\cite{genie}), with detailed interactions in low-$Z$ and high-$Z$ magnets and a generic geometry.
The Iron Core Magnets have been simulated by a massive cube of iron with a mass equivalent to the OPERA magnets and by using a 90\% inner
fiducial volume (770 tons and 330 tons for the far and near locations, respectively). Single hits have been extracted at the position of the RPC detectors and a 
realistic resolution has been included (0.75 cm for the single hit).
The ACM has been simulated as an empty 1 m deep region with 0.1 T magnetic field, limited by two vertical aluminum slabs. 
The overall simulation corresponds to the GLoBES software~\cite{globes} (version 3.0.11), to which it was partially compared and cross checked.

The muon reconstruction requires a minimal penetration length that corresponds to a 0.5 GeV cut. A conservative resolution 
of 10\% was used for the momentum evaluation, while the charge mis--identification was fixed at the 1\% level. 
Systematic effects due to mis--calibration of the detectors have been conservatively taken
at 0.5\% and 1\% levels, which include systematics due to the {\em relative} flux at the near and far positions.

\begin{figure}[htbp]
\begin{center}
  \includegraphics[width=0.68\textwidth]{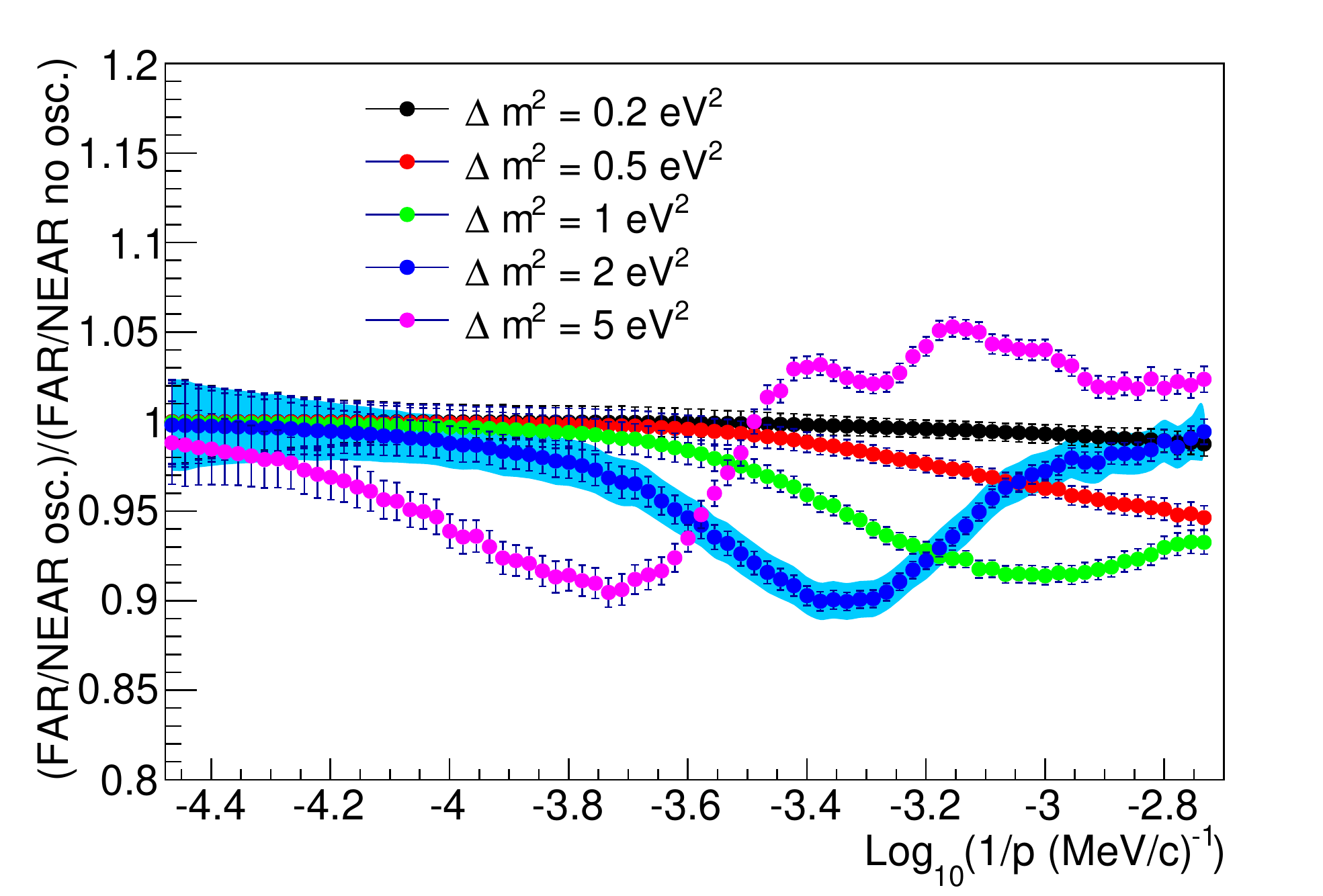}
    \caption{The double ratio of the disappearance probabilities at the near and far sites, 460 and 1600 m, respectively,    
    by using the amplitude 
    provided by the reactor anomaly, 0.146, and mass scales $\Delta m^2$ ranging from 0.1 to 5 eV$^2$ as a function of  $\log_{10}(1/p_\mu)$, where
     ``{\em p}'' is the reconstructed muon momentum in MeV/c. We apply cuts on the minimum muon path length (25 cm of range in iron) and
     the minimum momentum ($500$ MeV/c). A 90\% 
    fiducial acceptance of the magnet volume and an uncorrelated 1\% systematic error (from OPERA experience~\cite{opera-ele}) are also considered.
    Data collection corresponds to 1 year, i.e. $4.5\cdot 10^{19}$ p.o.t. The error bars correspond to the statistical errors. The blue band for $\Delta m^2=2$ eV$^2$
    is obtained by including the systematic error summed quadratically to the statistical one, the same result being obtained for the other mass scales.}
    \label{ster-4}
\end{center}
\end{figure}

In Fig.~\ref{ster-4} the double ratio {\em Far/Near} is shown as obtained by using the reconstructed muon momentum, for different 
values of $\Delta m^2$ scales. The relevant behaviour of the estimator is confirmed for $\Delta m^2$ values above $\approx 2$ eV$^2$
also in this very realistic and conservative approach. In fact, by using the muon reconstructed momentum in CC events instead of the true
 $\nu_{\mu}$ energy the oscillations properties will be affected 
by the kinematics and the interaction processes (it is worthwhile to note that there are mostly quasi--elastic interactions in the accounted--for energy range).

\begin{figure}[h]
\begin{center}
  \includegraphics[width=0.7\textwidth]{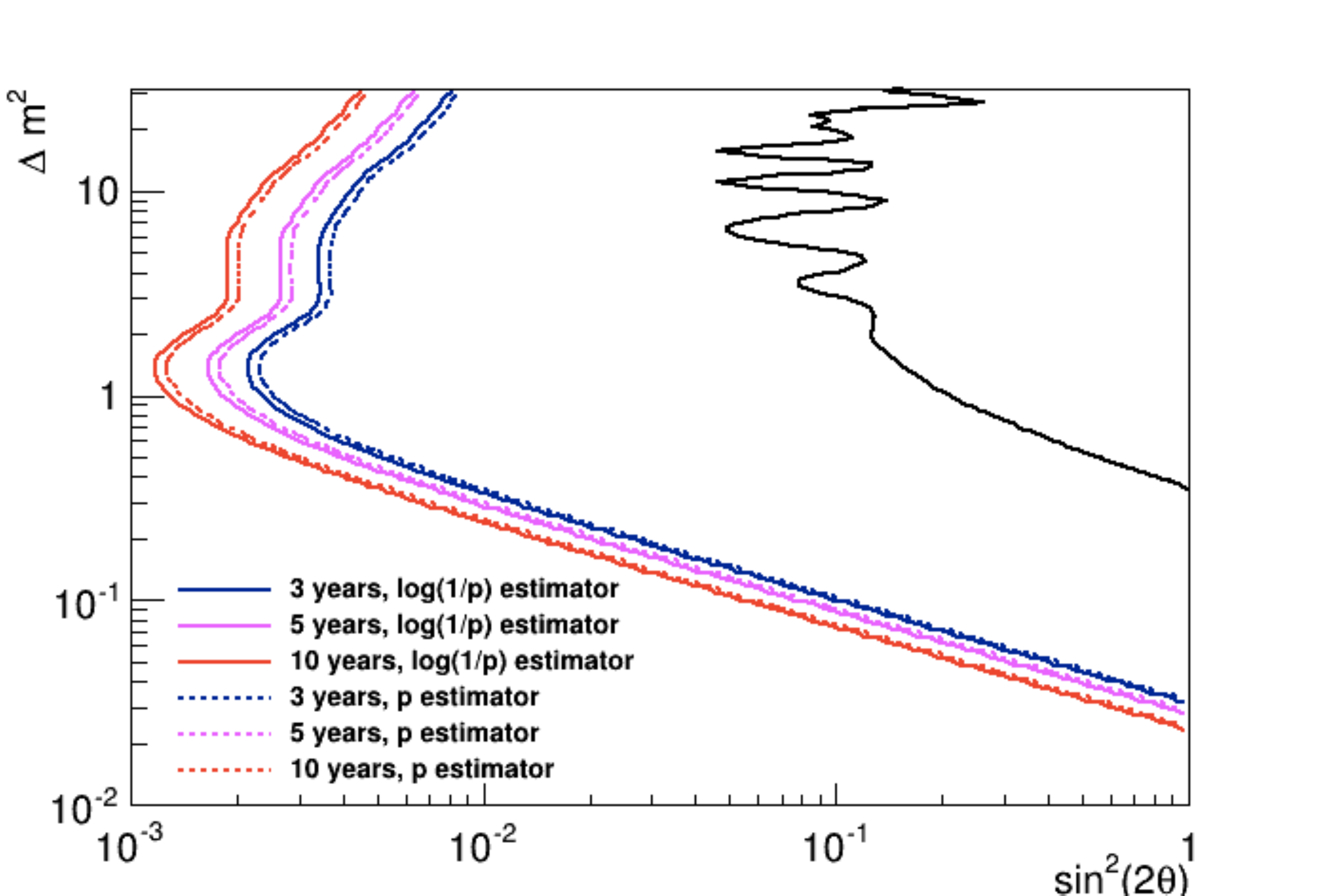}
   \includegraphics[width=0.7\textwidth]{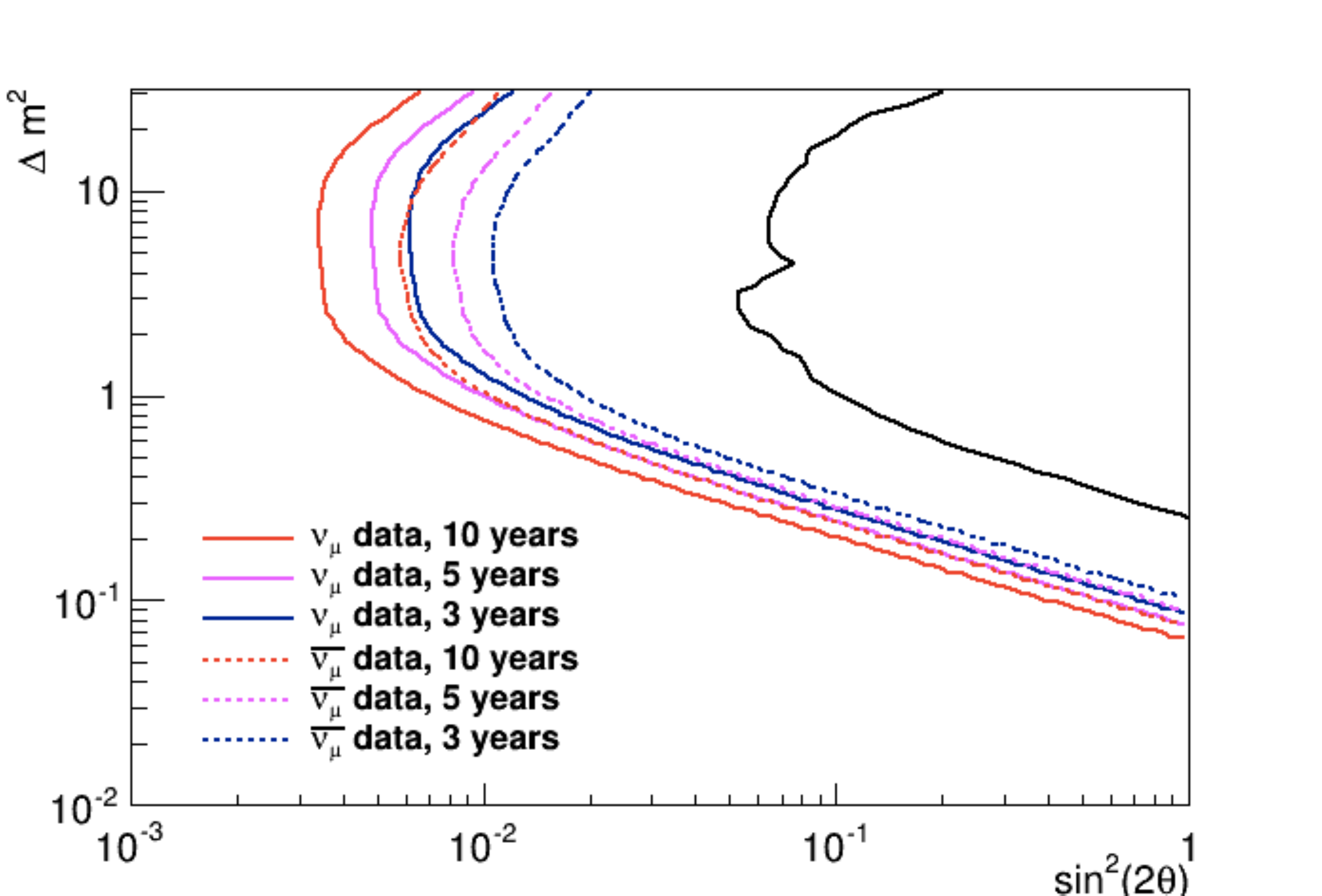}
    \caption{The estimated limits at 95\% C.L. for $\nu_{\mu}$ disappearance at a Short--Baseline beam at CERN for several luminosity running periods and different beam polarities,
    with a two--site massive spectrometer (770 tons and 330 tons, respectively) with 90\% inner fiducial volume.\\
    The top figure refers to the positive polarity beam. The continuous (dashed) lines correspond to the sensitivity limits obtained with the 
    $\log_{10}(1/p)$ ($p$) variable. 3 years correspond to $13.5\cdot 10^{19}$ p.o.t., 5 years to $22.5\cdot 10^{19}$ p.o.t. and 10 years to $45.0\cdot 10^{19}$ p.o.t.
    The exclusion limit from combined MiniBooNE and SciBooNE $\nu_{\mu}$ disappearance result at 90\% C.L. from Ref.~\cite{mini-sci-mu1} is shown for comparison by the black curve in the right.\\
   The bottom figure refers to the negative polarity beam. Sensitivity limits are evaluated with the $\log_{10}(1/p)$ variable.
     Clearly the negative polarity run allows the contemporary analysis of the  $\overline\nu_{\mu}$ and $\nu_{\mu}$ disappearance exclusion regions 
     thanks to the disentangling of the muon charge on an event--by--event--basis. The black curve in the right shows for comparison the central value of the sensitivity at 90\% C.L.  from
     combined MiniBooNE and SciBooNE $\overline\nu_{\mu}$ disappearance result (Ref.~\cite{mini-sci-mu2}).}
    \label{ster-5}
\end{center}
\end{figure}

Finally, in Fig.~\ref{ster-5} the estimated limits at 95\% C.L. on $\nu_{\mu}$ disappearance that can be achieved via the {\em Far/Near} estimator are shown for different data periods
(3, 5 and 10 years, corresponding to $13.5\cdot 10^{19}$, $22.5\cdot 10^{19}$ and $45.0\cdot 10^{19}$  p.o.t., respectively). The different results for $\nu_{\mu}$
and $\overline\nu_{\mu}$ beams were evaluated using the two variables, $p$ and $\log_{10}(1/p)$. In negative polarity runs the muon charge identification allows 
an independent, simultaneous and similar--sensitivity measurements of  the $\overline\nu_{\mu}$  and $\nu_{\mu}$ disappearance rates, due to the large $\nu_{\mu}$ 
contamination in the $\overline\nu_{\mu}$  beam. 

The parametrization used for the neutrino energy, $\log_{10}(1/E)$, which we conservatively prefer to address as $\log_{10}(1/p_{\mu-rec})$,
provides slightly better limits in almost all the ($\Delta m^2$, $\sin^2(2\theta)$) excluded regions.
This is partially due to the gaussian shape sensitivity of $1/p$ when cuts are applied to the corresponding variable. Moreover, the elucidation of the behaviour of the {\em Far/Near} ratio 
in terms of depletions and excesses in different regions of the spectra allows a better comparison between unoscillated and oscillated hypotheses.

As a result more than an order of magnitude improvement can be obtained in the sensitivity to the mixing parameter space between standard neutrino
and sterile ones with respect to today's limits over the whole $\Delta m^2$ range investigated. Specifically the mixing angle sensitivity (at 95\% C.L.) reaches $2.2\times 10^{-3}$ at 
$\Delta m^2 = 1$ eV$^2$, while at full mixing a $\Delta m^2=0.03$ eV$^2$ sensitivity at 95\% C.L. is obtained already with only 3 years of data collection with the neutrino beam.

Slightly worse results are obtained by collecting data with a negative polarity beam. The limiting mixing angle sensitivity at 95\% C.L. is around  $10^{-2}$ at 
$\Delta m^2 = 1$ eV$^2$. At full mixing a $\Delta m^2=0.1$ eV$^2$ sensitivity at 95\% C.L. is obtained with 3 years of data collection with the antineutrino beam,
the limiting factor being the intensity of the $\overline\nu_{\mu}$ beam. However it is worthwhile to note that by operating in negative polarity similar sensitivities can be obtained 
at the same time also for the neutrino component due to the large contamination of $\nu$ in the anti-$\nu$ beams.

\begin{figure}[h]
\begin{center}
  \includegraphics[width=0.7\textwidth]{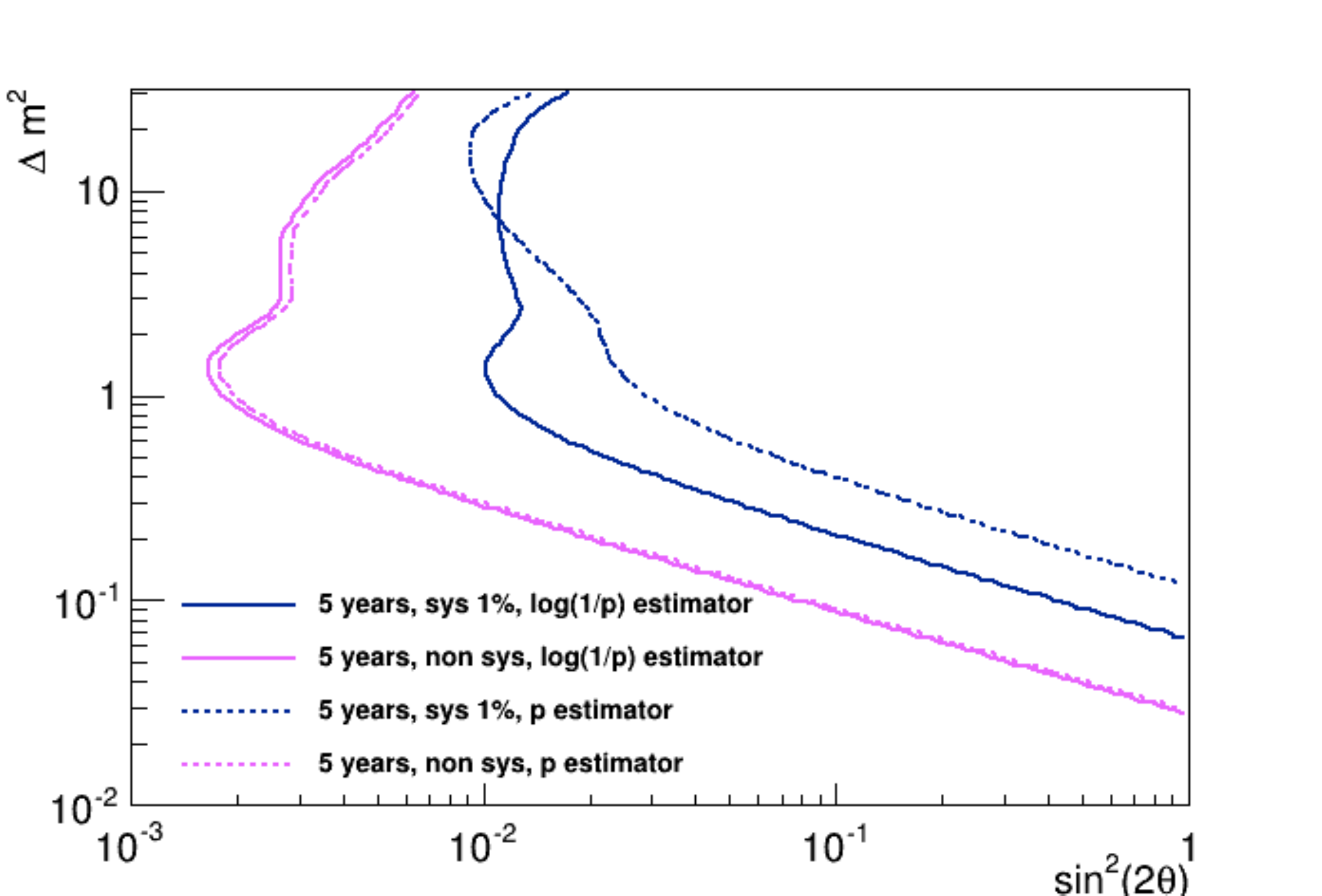}
   \includegraphics[width=0.7\textwidth]{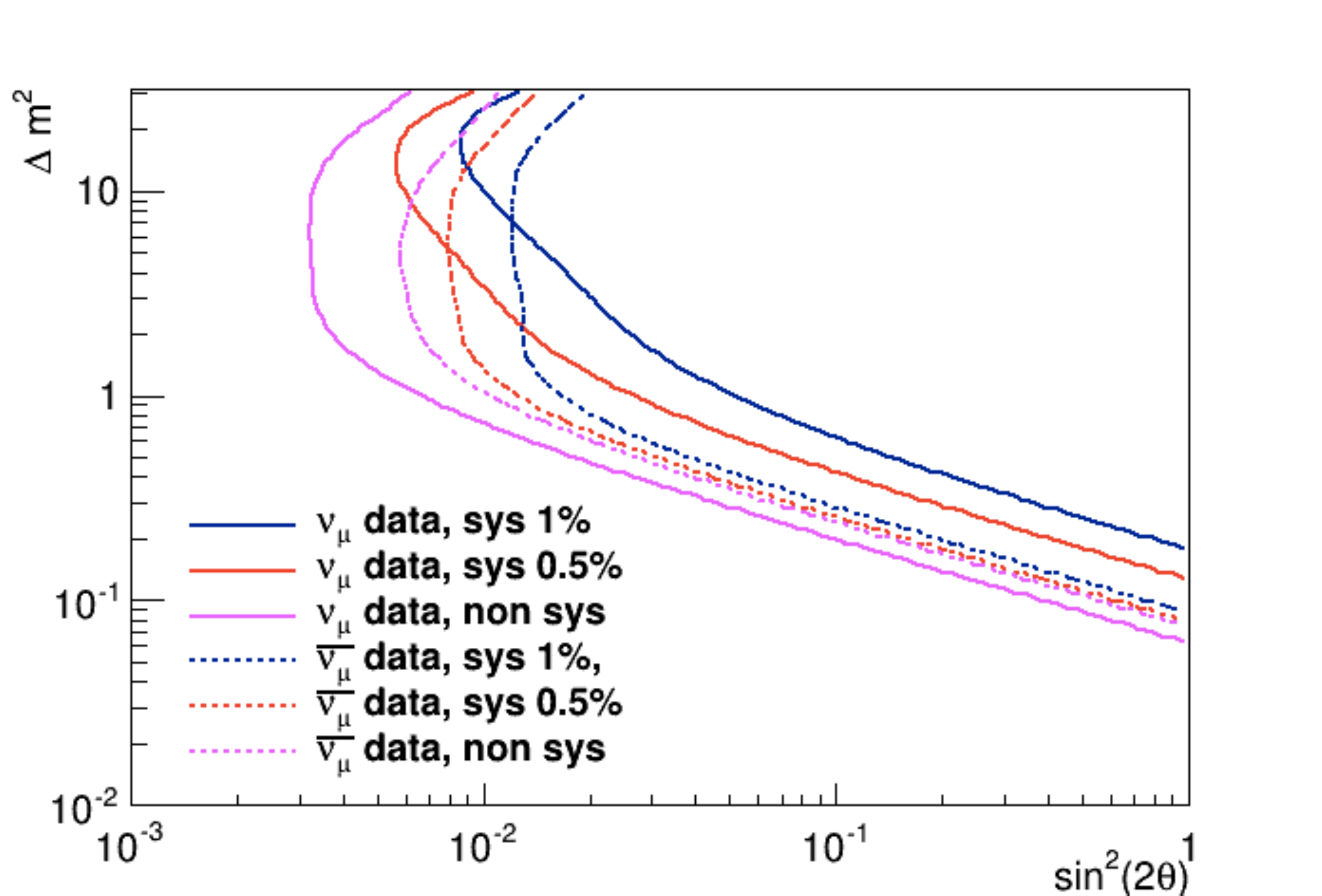}
    \caption{Systematic effects for the estimated limits at 95\% C.L. for $\nu_{\mu}$ disappearance at a Short--Baseline beam at CERN for 5 (positive polarity, top) and 10 (negative polarity, bottom) years of data taking,
    with a two--site massive spectrometer (770 tons and 330 tons, respectively) with 90\% inner fiducial volume. The exclusion regions are evaluated by using the $\log_{10}(1/p)$ (continuous lines)
    and $p$ (dashed lines) variables in case of the positive polarity beam data taking.}
    \label{ster-6}
\end{center}
\end{figure}

The relevance of statistics in presence of systematic effects is depicted in Figs.~\ref{ster-6}. It is evident that the overall error, dominated by the sensitivity 
on the measurement of the
muon momentum, puts an intrinsic  limitation to the data statistics which can be collected. As expected the useful luminosity that can be collected is asymptotically limited 
by systematics. The positive effect of the variable $\log_{10}(1/p_{\mu-rec})$ is more evident when larger errors are in place.

\section{Conclusions}

Neutrino physics is receiving more and more attention as a venue for the long standing search for
new physics beyond the Standard Model. The current anomalies which do not fit into the established standard scenario with 3 neutrinos
deserve refined studies and experiments. The CERN proposal for a new Short--Baseline experimental project is a very valuable one.
We illustrated the current critical tensions in the muon-neutrino disappearance field and the achievements that can be obtained
within the CERN project. Specifically, by considering $\mathcal{O}$(1) kton massive spectrometers,
 an improvement by an order of magnitude can be obtained in the sensitivity to the mixing parameter space between standard neutrinos
and sterile ones with respect to today's limits. Conversely, a possible $\nu_{\mu}$ disappearance signal will be essential to measure the
relevant physical parameters and to fully disentangle the different sterile models.

An effective analysis can be performed with a two-site experiment by using muon spectrometers with a low-$Z$ part that allows
clean charge identification on an event-by-event basis, and with a massive part allowing clean momentum  measurement through range.
Such a kind of spectrometer is under study by the NESSiE Collaboration and might be available with a limited investment.
The performances of these kinds of detectors, enlightened by the type of analysis developed in this paper, are suitable to put a
definitive result on the sterile neutrino issue at the eV mass scale.

\subsection*{Acknowledgements}

We are heartily dependent of the contributions of the NESSiE, ICARUS and CERN-CENF groups in developing the CERN project and the experimental proposals.
We wish also to warmly thank  the encouragements of Marzio Nessi and Carlo Rubbia in supporting such studies. We are finally in debit to Maury Goodman for
suggested criticisms and a careful proof--reading of the manuscript.

\clearpage
\newpage

%\newpage

%\vskip 30pt
%\newpage
%\centerline{\bf The NESSiE Collaborations}

%%\author{\noindent \\ \NessieAuthorList }

%\footnotesize{\noindent \\ \NessieAuthorList }

%\begin{flushleft}
%\footnotesize{\NessieInstitutes }
%\end{flushleft}

\end{document}